# Beyond structural stabilization of highly-textured AlN thin film: the role of chemical effects

O.V. Pshyk*, J. Patidar, S. Siol*

Empa – Swiss Federal Laboratories for Materials Science and Technology, 8600 Dübendorf, Switzerland

Corresponding authors:
Oleksandr.Pshyk@empa.ch; Sebastian.Siol@empa.ch

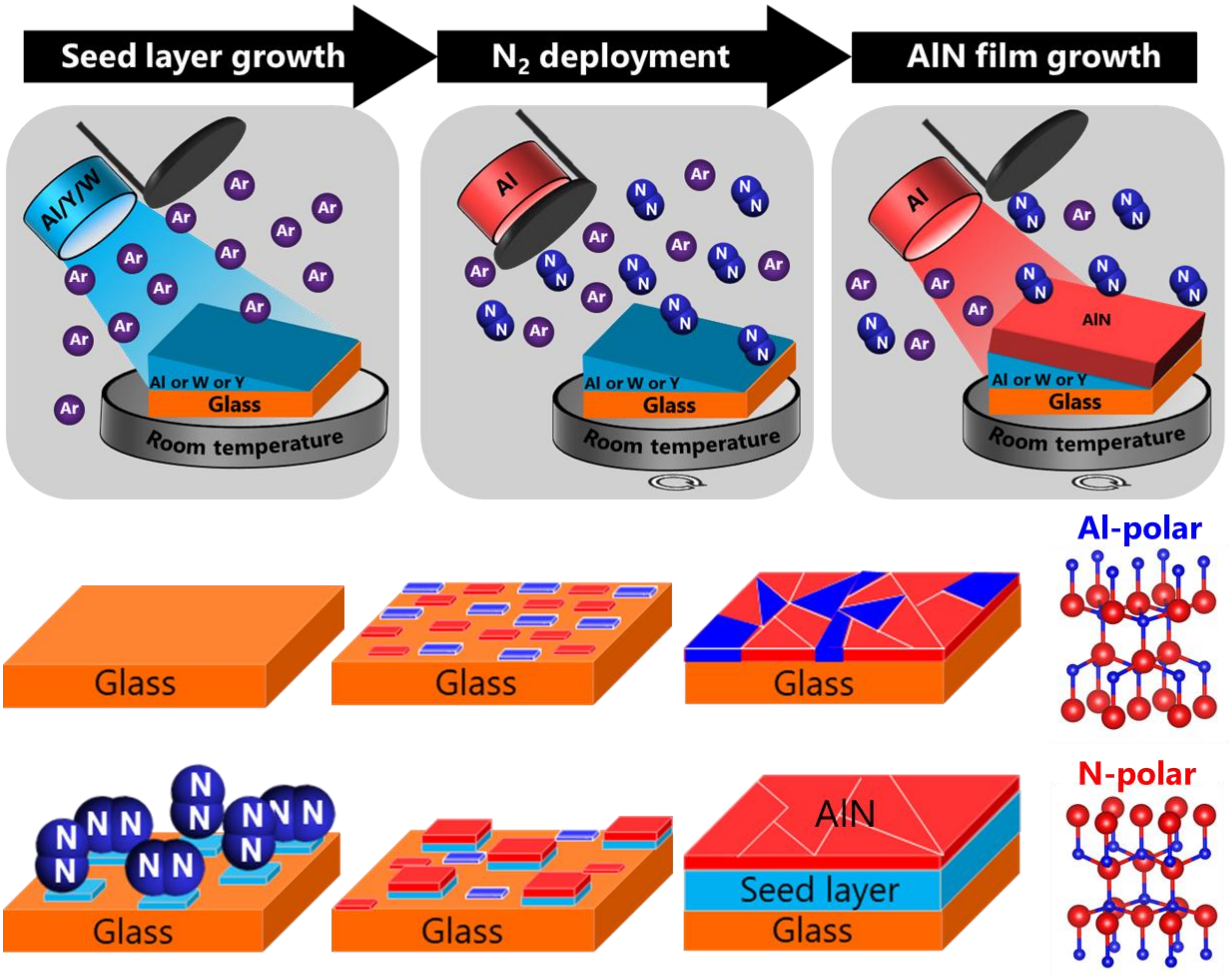

# Abstract

The crystalline quality and degree of c-axis orientation of hexagonal AlN thin films correlate directly with their functional properties. Therefore, achieving AlN thin films of high crystalline quality and texture is of extraordinary importance for many applications, but in particular in electronic devices. Here, we present a systematic study revealing that the growth of c-axis orientated AlN thin films can be governed by a chemical stabilization effect in addition to the conventionally known structural, i.e. strain-induced, stabilization mechanism. The promotion of in-plane growth of AlN grains with c-axis out-of-plane orientation is demonstrated on Y, W or Al seed layers with different thicknesses and crystallinity preliminary exposed to $N_2$ at room temperature. We establish that the stabilization mechanism is chemical in nature: the formation of an N-rich surface layer on the metal seed layers upon exposure to $N_2$ pre-determines the polarity of AlN islands at initial stages of thin film growth while the low energy barrier for the subsequent coalescence of islands of the same polarity contributes to grain growth. These results suggest that the growth of c-axis oriented AlN thin films can be optimized and controlled chemically thus opening more pathways for energy-efficient and controllable AlN thin film growth processes.

# Introduction

Hexagonal AlN thin films have found a broad range of applications such as surface acoustic wave (SAW) devices, film bulk acoustic resonators (FBAR) or microelectromechanical systems (MEMS) due to their high acoustic velocity, chemical resistance, thermal stability, large dielectric breakdown field and linear frequency response[1–4]. In fact, high-quality AlN thin films are essential building blocks in many devices we use every day[5]. However, AlN thin films demonstrate the best performance only when grown with a high quality, i.e. large crystals with low defect density and pronounced c-axis orientation[3].

Achieving high-quality of AlN thin films on metal substrates or underlying layers is of great importance, especially for MEMS and FBAR devices. A variety of deposition parameters can be tuned to improve the crystallinity and texture of AlN films[6]. The most common approach is the combination of a suitable substrate with a relatively high substrate temperature. Alternatively, controlled low-energy ion bombardment has been shown to improve the crystallinity of AlN thin films [7–9]. However, AlN films grown at high temperatures can suffer from internal residual stresses due to the thermal expansion coefficient difference and lattice mismatch between the AlN and the underlying substrate. Similarly, low-energy ion bombardment can lead to process gas incorporation and high residual compressive stress. This can cause high defect density in AlN thin films leading to excessive deformation, generations of cracks, and eventual film delamination that deteriorates the device's performance. Moreover, high growth temperatures are energy-intensive and incompatible with a number of device layouts and substrates. The limited choice of suitable substrates for the growth of high-quality AlN films satisfying requirements for a given application has been overcome by using seed metal layers. A wide variety of metal seed layers have been studied for the growth of AlN films. Pt[10] and Mo[1] are among the best AlN growth promoters due to a structural, i.e. strain-induced, stabilization mechanism. On the other hand, a set of case studies shows that even a minute change in the substrate composition can have a tremendous effect on AlN thin film growth evolution, implying the existence of a stabilization mechanism beyond such structural effects. Yet, the results are not fully understood and warrant further investigation[11–13].

The decisive role of the structural relationship between the AlN thin film and the substrate originates from the necessity to promote a layer-by-layer two-dimensional thin-film growth, facilitating epitaxial growth or local epitaxy at the length scale of individual grains of a given orientation. This growth mode is driven by specific substrate/film interface interactions defined by a minimum in the total energy of the film-substrate system. The selection of an appropriate substrate or seed layer can provide low interface energy and low elastic strain energy through coherent interfaces and low lattice mismatch with AlN. This is decisive for the growth of highly-textured thin-films with large grain sizes.

However, chemical contribution to the interface energy due to the formation of mismatching chemical bonds at the interface can have another important consequences for the initial stages of thin-film growth and subsequent microstructural evolution[14,15]. Indeed, a few early studies show that substrate surface composition and chemical nature, in addition to its structure, can have a tremendous effect on AlN thin-film morphology and the associated performance[11–13,16–18]. For example, the deposition of an ultra-thin Al layer (~0.2 nm) on Si substrates before AlN growth at high temperature significantly improves the quality of films[12] while Ru surface decoration with $O_2$ was used to tune the polarity of AlN thin films[11]. Moreover, under non-equilibrium growth conditions, any minor additives on the seed layer surface can significantly change the potential energy landscape, and thus might impact energy barriers for adatom diffusion leading to the modification of growth kinetics and the growth mode. Despite the widespread application of metallic seed layers for the growth of AlN thin films, systematic studies of other mechanisms beyond structural templating, which can be responsible for high-quality AlN thin film growth, to the best of our knowledge, are not reported. Therefore a deeper understanding of AlN thin-film growth mechanisms on metallic seed layers is required to fully leverage this deposition strategy. In turn this would enable the deposition of high-quality thin films at relatively low temperatures, for a wider and more efficient technological applicability of this material.

Here, we present a systematic study of AlN thin film growth on chemically and structurally different metallic seed layers, which are exposed to $N_2$ at room temperature prior to AlN deposition. The investigation focuses on magnetron sputter-deposited combinatorial metal seed layer libraries of W, Y and Al grown on borosilicate glass substrates for the subsequent growth of AlN thin films without breaking vacuum. We design and implement a deposition protocol to differentiate between either a chemical or a structural templating effect: we strategically vary the metal-layer thickness across the substrate from ~115 nm down to less than ~1 nm to tune the crystallinity of seed layers and their substrate coverage. Due to thickness-dependent crystallization, all seed layers grow with amorphous structure below a certain thickness that allows us to study chemical effects. The results of this work reveal a chemical stabilization mechanism rationalized in terms of the termination of metal seed layers surface with $N_2$ that promotes the growth of textured AlN thin films in a certain c-axis polarization.

# Experimental

Deposition of thin films is performed in an AJA 1500-F sputtering system pumped down to a base pressure lower than $10^{-6}$ Pa. All films are grown onto 50.8 × 50.8 mm$^2$ large and 1.1 mm thick borosilicate glass (EXG) substrates. The substrates are ultrasonically cleaned in acetone, and ethanol, and subsequently dried in a flow of $N_2$. Al, Y and, W seed layers with different thicknesses are grown by oblique angle radio-frequency magnetron sputtering (RF) from the corresponding metal targets of high purity (99.99 % purity each) and 50.8 mm in diameter. The sputter power densities are between 1.23 W/cm$^2$ and 2.45 W/cm$^2$. The unbalanced magnetrons are arranged in a confocal, sputter-up geometry with open-field magnetic configuration and a deposition angle of approximately 30° and a working distance of 10.5 cm. The deposition of seed layers is carried out at a pressure of 0.33 Pa at an Ar flow of 15 sccm. The seed layer depositions are performed without substrate rotation leading to well-controlled thickness gradients. Prior to AlN thin film growth, as-deposited seed layers are exposed to Ar+$N_2$ gas flow (8 sccm and 16 sccm for Ar and $N_2$, respectively) at a pressure of 0.5 Pa for 1 min. Immediately after that, 50 nm thick AlN thin films are grown by direct current magnetron sputtering of Al target in a mixed atmosphere of Ar (8 sccm) and $N_2$ (16 sccm) at a pressure of 0.5 Pa. For this deposition, the substrate is rotated to ensure a uniform film across the entire substrate. The deposition sequence used in this work is summarized in Fig. 1. All steps are performed without intentional substrate heating in order to eliminate effects of temperature during $N_2$ exposure of the seed layers (step b).

The structural properties of the films are measured using a Bruker D8 X-ray diffraction (XRD) system equipped with a Cu Kα X-ray source and operating in Bragg-Brentano geometry. The full width at half maximum (FWHM) of the selected reflections is determined by peak fitting using a pseudo-Voigt function. The FWHM is used for calculation of the average grain size of the seed layers using the Scherrer equation, $FWHM_{hkl} = K\lambda/D \cos\Theta_{hkl}$, where K is a geometrical constant (taken as 1 for both cubic and hexagonal crystals assuming spherical shapes of the grains), $\lambda$ is the X-ray wavelength, D is the average grain size, $\Theta_{hkl}$ is half the Bragg angle 2Θ. The thickness of the seed layers is measured by a surface profilometer (Bruker Dektak XT).

For an investigation of the seed layer's surface prior to the AlN deposition *in-situ* X-ray photoelectron spectroscopy (XPS) analysis was performed. Immediately after step (b) of the deposition process (Fig. 1) samples with 50 nm thick Al or W seed layers on glass substrates are transferred from the deposition chamber to the spectrometer using a UHV transfer chamber. The samples are stored at a pressure below $10^{-4}$ Pa throughout the transport process. XPS measurements are performed using a Physical Electronics Quantera II Hybrid spectrometer using monochromated Al-Kα radiation. The base pressure during the spectra acquisition is below $10^{-6}$ Pa. Charge neutralization is achieved using a dual beam charge neutralization system based on a low-energy electron flood gun and a low energy positive ion source. The binding energy scale is calibrated by examining sputter-cleaned Au, Ag, and Cu reference samples according to the recommended ISO standards for monochromatic Al-Kα sources that place Au $4f_{7/2}$,Ag$3d_{5/2}$, and Cu $2p_{3/2}$ peaks at 83.96, 368.21, and 932.62 eV respectively[19,20]. For static charge correction, the energy scale is aligned based on ΔE value determined as a difference between Al 2p and W 4f main metal peak positions measured on Al and W thin films on glass substrate, respectively, and the binding energy of Al 2p and W 4f core-electrons reported by C. Powell[21] for pure Al and W, taken as 72.8 eV and 31.3 eV, respectively. XPS spectra quantification and fitting is performed using CasaXPS software package. Al 2p and W 4f spectra are fitted using constraints for the area ratio of spin-orbit split doublets as 0.5 and 0.75 for Al 2p and W 4f, respectively. Spectra fitting is performed using Voigt function following Shirley-background subtraction. Metallic components in Al 2p and W 4f spectra and W-N components in W 4f spectrum are fitted with asymmetrical line shapes in the form of Voigt functions with a tail modifier. No constrains are used for FWHM and the binding energy. The inelastic mean free path (λ) is calculated from the kinetic energy of the detected electrons based on the Tanuma, Powell, and Penn formula[22]. The modified Auger parameter (AP) is calculated as the sum of kinetic energy (KE) of a sharp core-core-core level type Auger transition (KE (XYZ)) and binding energy (BE) of the core-level photoelectron involved in the Auger transition (BE (X/Y/Z)), i.e. $\alpha' = KE\ (XYZ) + BE\ (X/Y/Z)$[23,24].

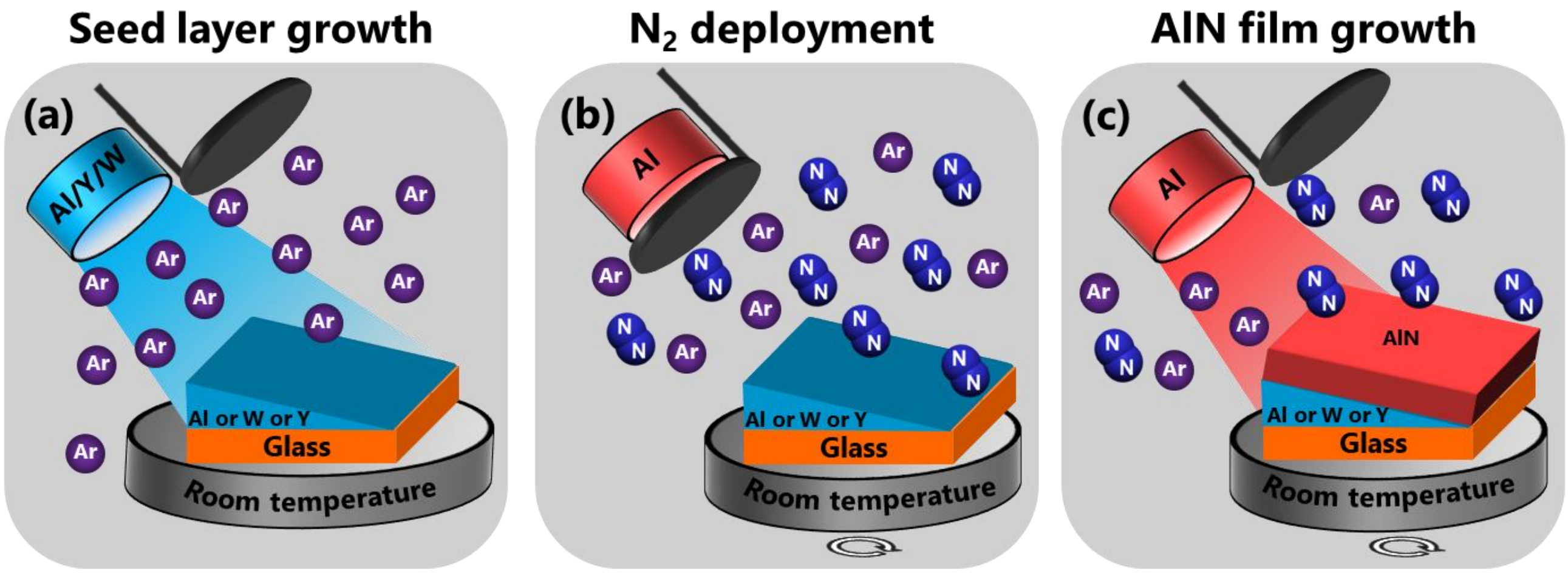

**Fig. 1.** Schematic representation of the design and implementation of the deposition sequence: (a) oblique-angle RF magnetron-sputter deposition of metal seed layers on multiple glass substrates resulting in sample libraries with thickness gradient across the substrate, (b) pre-conditioning of Al target in and exposure of the seed layers to Ar+$N_2$ gas flow (the shutter of the Al gun remains closed, substrate rotation is on), (c) DC magnetron-sputter deposition of 50 nm AlN thin film without intentional heating (substrate rotation is on).

# Results and discussion

AlN thin films of 50 nm thickness are grown on combinatorial metal seed layer libraries with different thicknesses to investigate the influence of the seed layers crystallinity and glass substrate coverage on the growth of AlN thin films. The thickness was varied in a range from ~133 nm down to less than ~1 nm. Representative XRD patterns for three extreme cases (no seed layer, amorphous seed layer and crystalline seed layer) for each seed layer material show that all three investigated seed layers have a positive effect on the crystalline quality of AlN thin films. This can be seen by comparing the AlN (0002) intensity: all films exhibit a pronounced out-of-plane texture along the [0002] direction as evidenced by no signal from any other AlN reflection for Al and W seed layers and only a minor signal from the [$10\bar{1}1$] orientation for Y seed layer. In contrast, XRD pattern from AlN thin films grown on the bare glass substrate shows a weak and broad reflection from the (0002) planes of AlN. Fig. 2b exhibits the full-width at half maximum (FWHM) of the AlN (0002) peak (further in the text AlN-FWHM), representative of grain growth, plotted as a function of Al, W, and Y seed layers thicknesses for the complete thickness libraries. The growth of [0001]-oriented AlN grains is promoted even upon a minute decoration of the glass substrate by metals while the largest AlN-FWHM is measured for the films on the bare glass substrate. Fig .1c shows the AlN-FWHM plotted as a function of Al, W, and Y seed layers grain size determined using Scherrer equation for the entire metal seed layer thickness-libraries. It is revealed that irrespective of the seed layer crystallinity or crystal lattice structure, AlN crystallization is improved on both X-ray amorphous (broad XRD peak or absence of peaks from seed layer material, e.g. Fig. 1a) and crystalline metal seed layers in comparison to the bare glass substrate.

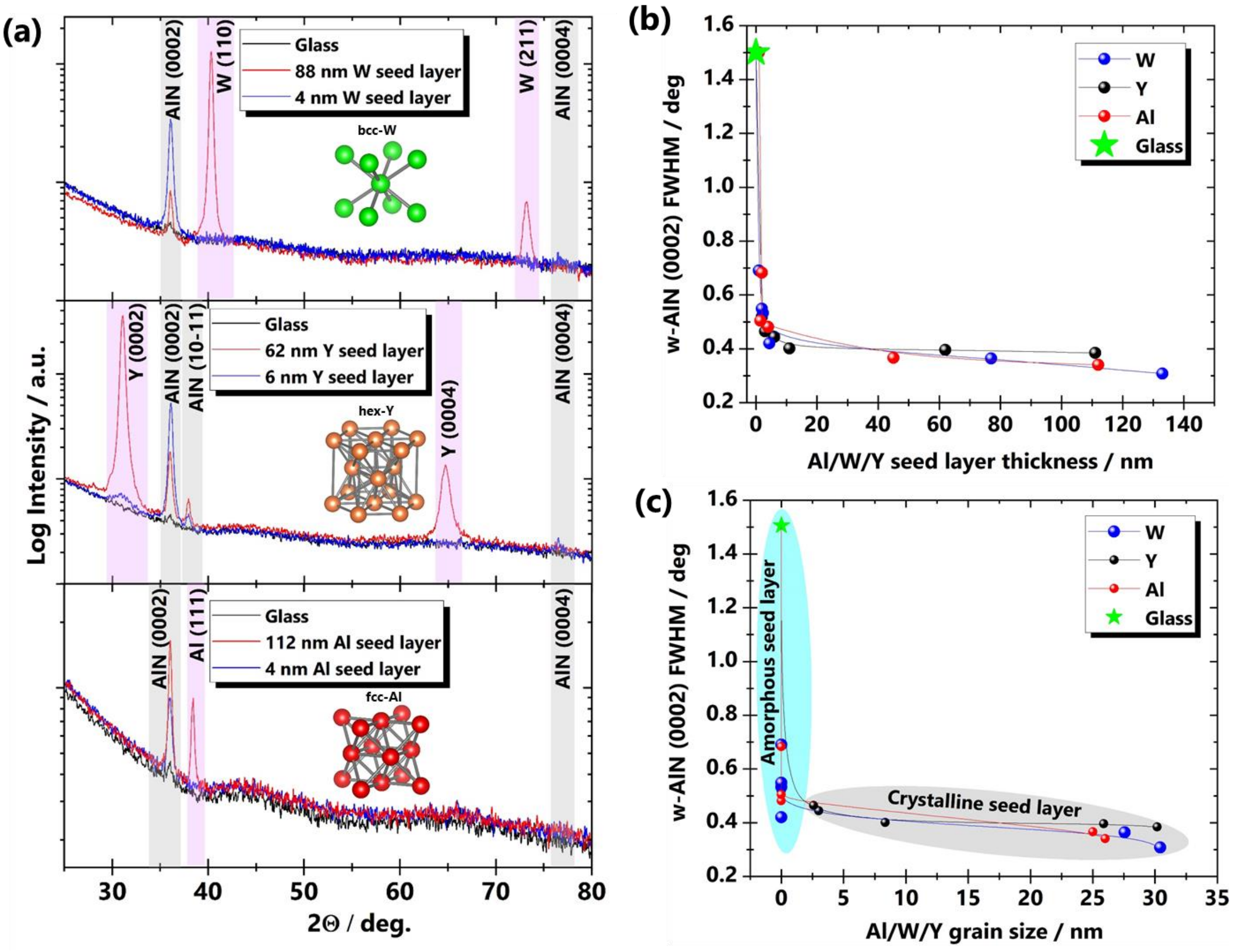

**Fig. 2.** (a) Representative XRD patterns from AlN thin films grown on W, Y, and Al seed layers of different thickness and crystallinity (the thinnest and thickest layers correspond to X-ray amorphous and crystalline seed layers, respectively), and XRD pattern from AlN thin films on bare glass substrate as a reference. (b) FWHM of the AlN (0002) peak plotted as a function of Al, W, and Y seed layers for the complete set of thickness libraries, (c) FWHM of the AlN (0002) peak plotted as a function of Al, W, and Y seed layer grain size determined using Scherrer equation. Crystal structures of body-centred cubic (*bcc*) W, hexagonal (*hex*) Y and face-centred cubic (*fcc*) Al lattices are show in (a) as insets.

Since both crystalline and amorphous seed layers equally promote AlN thin film growth, the mechanism lying behind the promotion seems to be structure-agnostic. Moreover, the glass substrate is also amorphous but AlN films grown on it demonstrate the largest FWHM. Consequently, the nucleation of AlN needs to be governed by a chemical stabilization mechanism. This motivates a study of the surface chemistry of the seed layers prior to the deposition of AlN thin films. In order to probe the surface chemistry of the seed layers *in-situ* XPS analysis was performed by UHV transfer of Al and W seed layers to the XPS chamber after a "dummy" deposition process stopped right before the shutter of magnetron was opened for AlN growth. This corresponds to the time right after step (b) in Fig 1. Fig. 3 shows the XPS analysis of the W and Al seed layers after this short exposure to the reactive process gas. Al-N chemical bonds are revealed on the Al seed layer surface evident as an Al $2p_{3/2}$ peak component shifted by 1.35 eV to higher binding energy relative to the main metal Al $2p_{3/2}$ peak component at 72.8 eV (Fig. 3a). The corresponding N 1s core-level (Fig. 3b) is fitted with three components. The peak at around 397.3 eV can be assigned to Al-N bonds because this binding energy is very close to that of N 1s electrons in AlN thin films[25,26]. The N 1s peak component at around 398.9 eV can be assigned to chemisorbed $N_2$ whereas the peak at 401.9 eV is due to N-O species. Similarly, W-N bonds on the W seed layer are evident as a component of the W $4f_{7/2}$ core-level emission (Fig. 3c) shifted by 0.59 eV to higher binding energy relative to the main metal component at 31.3 eV. The N 1s component at 397.4 eV is close to the binding energy of N 1s electrons in cubic $W_2N$[26] and therefore can be assigned to W-N bonds. Moreover, the component at 398.6 eV can be assigned to the chemisorbed $N_2$[27,28]. The intensity of Me-N (Me stands for Al or W) components of the metal core-level spectra is relatively low in comparison to the main metal component because Me-N bonds reside at the outermost surface while the probing depth of the most of detected photoelectrons is much beyond the thickness of Me-N layer. Importantly, the peak area of the metal nitride and chemisorbed $N_2$ peaks in the N 1s core-level spectrum $[I(\text{Me-N}_{N1s})+I(\text{N2}_{N1s})]/\text{RSF}_{N1s}$ is equal to the metal nitride peak area in metal core-level spectrum $I(\text{Me-N}_{Al2p/W4f})/\text{RSF}_{Al2p/W4f}$, where $I$ is the peak area and RSF is the relative sensitivity factor. This implies that the binding energy of Al 2p and W 4f electrons from Al and W atoms chemically bonded to chemisorbed $N_2$ is close to or equal to that of Al and W atoms bonded to N in the near-surface MeN layer. The depth-integrated peak area ratio $I$(Me-Me)/$I$(Me-N) in the spectra of metal core-levels, where $I$(Me-Me) is the metal peak area and $I$(Me-N) is the metal nitride peak area, determined considering an exponential attenuation of the original signal as following:

$$\frac{I(\text{Me-N})}{I(\text{Me-Me})} = \frac{\int_0^d e^{-d/\lambda}\,\partial d}{\int_d^\infty e^{-d/\lambda}\,\partial d},$$

where *d* is the thickness of the MeN surface layer, λ stands for the inelastic electron mean free path of electrons in Al or W ($\lambda_W$ = 1.984 nm and $\lambda_{Al}$ = 3.280 nm), is used to estimate the thickness of the MeN surface layer formed on Al and W seed layers upon exposure to $N_2$. The estimated nitride layer thickness is 0.1 nm and 0.25 nm on Al and W seed layers, respectively.

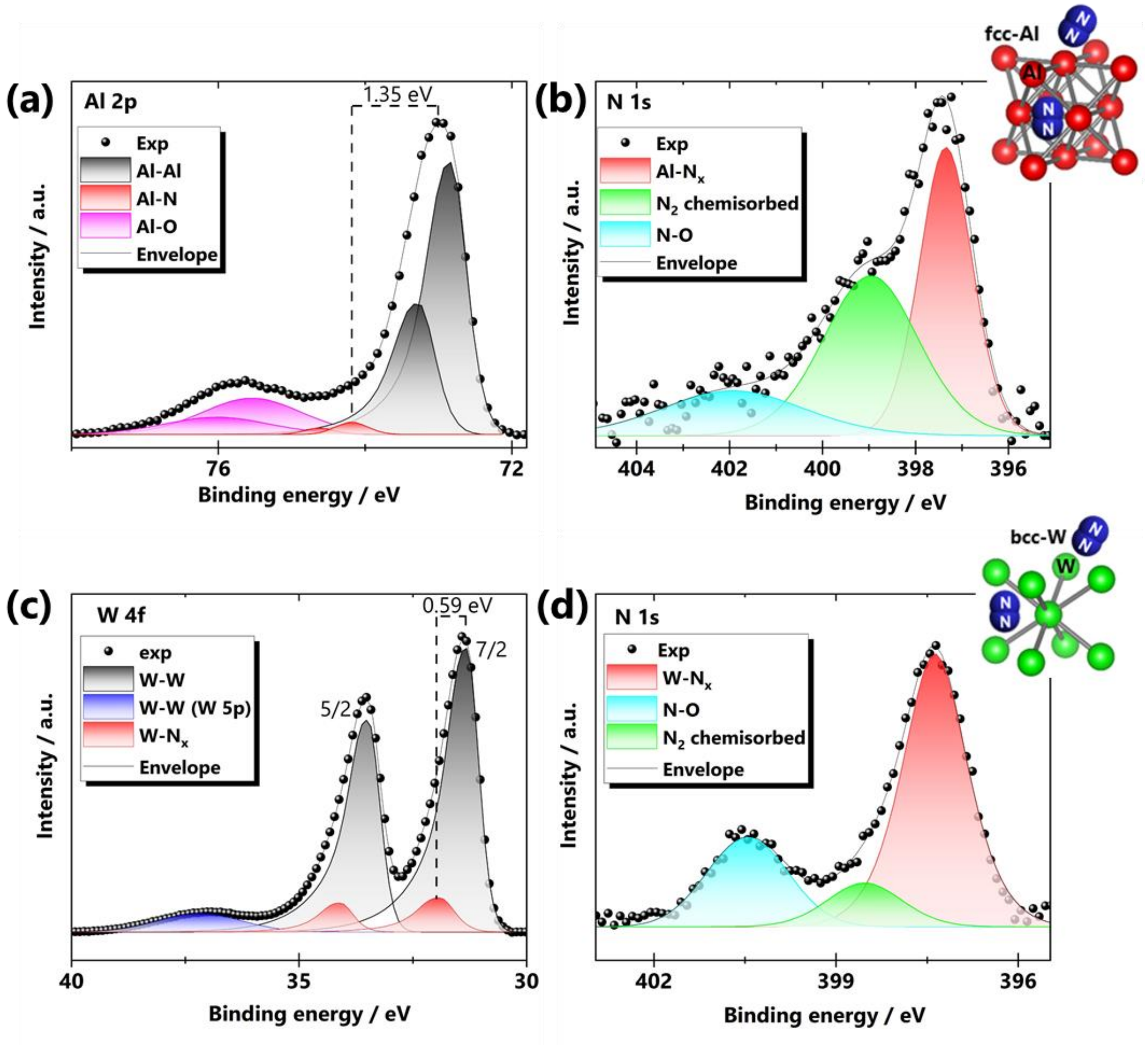

**Fig. 3.** UHV transfer XPS analysis for Al and W seed layers exposed to $N_2$/Ar flow (see Fig. 1b) prior to AlN growth: (a) Al 2p and (b) N 1s core-level spectra acquired from Al seed layer, and (c) W 4f and (d) N 1s core-level spectra acquired from W seed layer.

Auger electrons have a low probing depth due to their low kinetic energy, which makes them more sensitive to the changes in the surface composition than photoelectrons. In addition, Auger parameter analysis is less sensitive to surface charging effects.[29,30] Therefore, Auger parameter (AP) analysis is performed to gain more insights about chemical properties of the outer layer of metal seed layers after exposure to $N_2$. The N AP is calculated using the maxima of the N $KL_{2,3}L_{2,3}$ peak (Fig.

4a, b) and the respective components of the N 1s peak for both Al and W seed layers. The results are summarized in the form of a Wagner plot (Fig. 4c) together with AP of the relevant reference metal nitrides for comparison. Fig. 4c shows a pronounced shift in the AP reflecting a high sensitivity of N AP to changes in the local chemical environment around N atom. The N AP is the largest for the chemisorbed $N_2$ residing on the outer surface while N AP for N from the MeN layer shifts towards the values of corresponding metal nitrides. In addition to the fact that the binding energy of the main component of N 1s is very close to the binding energy of N 1s photoelectron in metal nitrides, the AP study confirms that coordination and chemical bonding of N in the MeN surface layer is close to that in corresponding metal nitrides because the corresponding AP locates between the AP of metal nitrides and the chemisorbed nitrogen. Concluding, N resides on the surface of the metal seed layers in two different chemical states prior to AlN thin film growth: chemisorbed $N_2$ and N constituting an ultrathin MeN surface layer.

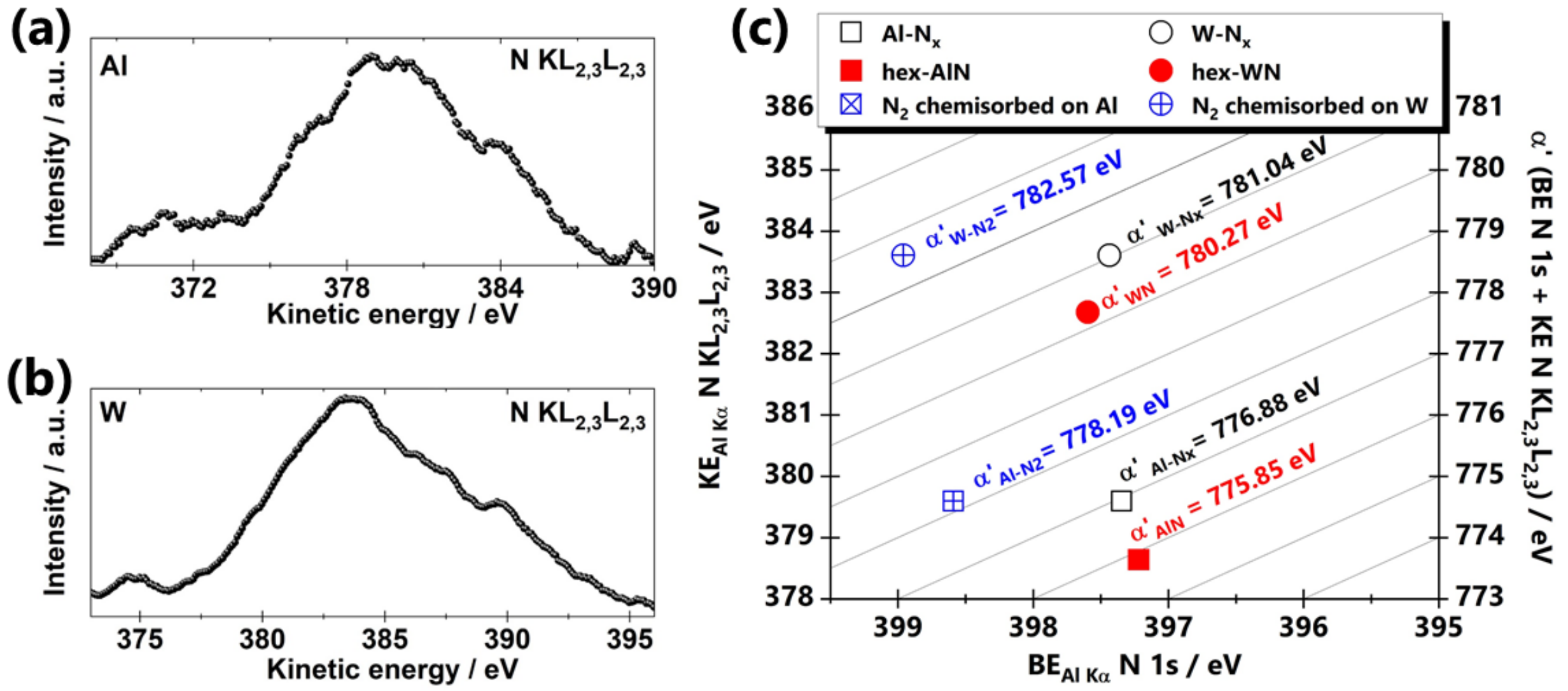

**Fig. 4.** N $KL_{2,3}L_{2,3}$ Auger emission lines of Al (a) and W (b) seed layers exposed to $N_2$/Ar gas for 1 min and Wagner plot for Auger parameter of N 1s core-levels and N $KL_{2,3}L_{2,3}$ Auger emission lines. N AP in polycrystalline hexagonal AlN (*hex*-AlN) and WN (*hex*-WN) thin films are shown as references (un-reported studies).

It is apparent that metal seed layer surfaces are terminated with a N-rich surface layer before the growth of AlN thin films. In stark contrast, the glass substrate is unaltered upon exposure to $N_2$. This principal difference becomes decisive for the subsequent nucleation and growth of AlN thin films. A schematic image summarizing different mechanisms of AlN thin film growth on glass substrate and seed metal layers is presented in Fig. 5. It is well known that AlN thin film grow N-polar or Al-polar, i.e. with $[000\overline{1}]$ or [0001] preferential orientation, respectively, on metal and ceramic crystalline seed layers depending on the surface termination[11,31,32]. Moreover, it is well known from the previous studies that the same growth mechanisms leading to a more pronounced [0001] texture and crystalline quality lead also to a preferential growth of one of the polarizations[4]. Typically in reactive sputter deposition of AlN, the predominant polarity is N-polar, especially when the growth is performed on metallic seed layers.[32–34] However, during nucleation competition of islands with different polarity can influence grain growth. The overall growth evolution of AlN thin films presented in this work suggests that the amorphous glass substrate does not act as a growth promoting template to none of the two possible polarizations of AlN grains leading to competitive grain growth and eventual small grain sizes. The competitive grain growth in this scenario occurs between islands of opposite polarities because the energy penalty for the coalescence of two islands with opposite polarities is large. Since the pristine metal seed layer surface exposed to $N_2$ develops a MeN sub/monolayer with N at the outermost surface. The incident Al atoms of the sputter flux preferentially bond to the outermost N atoms forming thus the first islands of the growing AlN film. Therefore the AlN grains grow Al first. Next, the incident N atoms of the sputter flux react with Al, forming the next monolayer and so on leading to AlN thin film growth. Such nucleation path pre-determines the polarity of the islands and further growth evolution leading to AlN growth in this pre-determined polarization. Therefore, even the deposition of an ultra-thin amorphous metal seed layer and its exposure to $N_2$ significantly alters the evolution of AlN thin film growth; AlN islands of a given polarization win in competitive growth due to a relatively low energy penalty for the coalescence of the islands of the same polarity leading to increment of the grain size irrespective of seed layer crystallinity. Therefore, AlN thin film growth is promoted even on amorphous seed metal layers (Fig. 2) where strain-induced stabilization is not accessible. In addition, a homogenous N-passivated surface might increase the diffusion-length of adatoms further promoting grain growth.

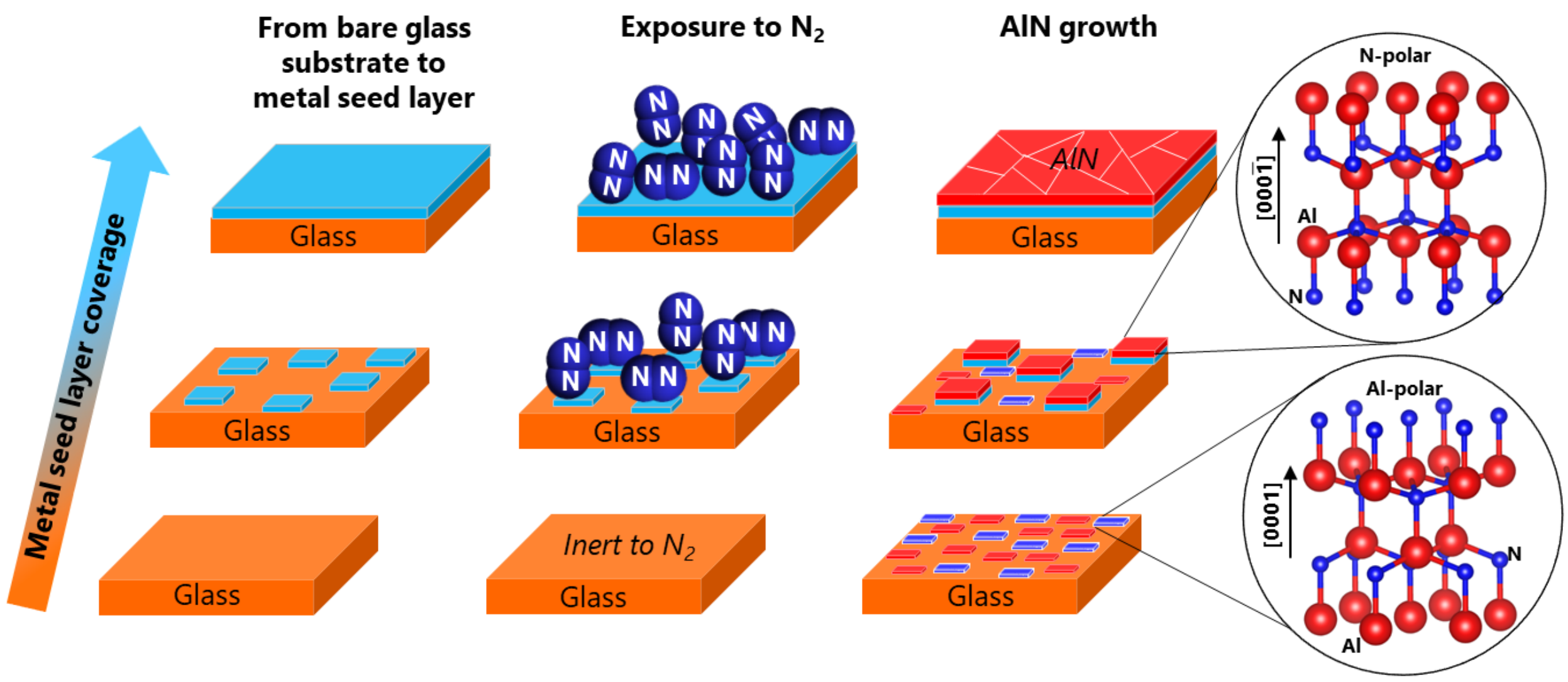

**Fig. 5.** The schematic of AlN thin film growth mechanism: from bare glass substrate to a metal seed layer terminated with N: glass substrate is inert to $N_2$ in contrast to metal seed layers that results in the formation of an N-rich surface layer on the later after exposure to $N_2$ before AlN growth. Competitive growth of AlN grains with two polarities on the bare amorphous glass substrate changes to the promotion of AlN grain growth with pre-determined polarization on metal seed layers.

## Conclusion

In summary, this study unveils an alternative approach to the conventional strategy for stabilizing and promoting the growth of AlN nitride thin films with a specific orientation and a large average grain size. Alongside the optimal selection of a substrate based on lattice-mismatch and symmetry considerations, we demonstrate that AlN grain growth can be further influenced and modified by chemical effects. Through *in situ* XPS on metal seed layers exposed to $N_2$ prior to AlN growth, we reveal the formation of a N-rich surface layer on metal seed layers. Considering that AlN thin film growth is equally promoted on X-ray amorphous and crystalline metal seed layers upon exposure to $N_2$, the promotion can be univocally assigned to a chemical effect only set by N surface termination. We propose that this is based on the preferential bonding of Al from the sputter flux to N-rich surface layer resulting in the nucleation of AlN islands with predefined polarity at the initial stage of film growth and low energy barrier for their further coalescence, leading to grain growth. The mechanism uncovered in this study expands the fundamental understanding of the AlN growth process on various metallic substrates, complementing the conventional strain-driven mechanism and AlN/metal interface symmetry considerations. The newfound insight is poised to facilitate the deposition of high-quality AlN thin films with pre-defined polarity at lower temperatures on a variety of different substrates. The revealed mechanism introduces additional avenues for controlling and optimizing the growth of AlN thin film, promoting the development of more energy-efficient deposition processes and enhancing process flexibility for the design of AlN-based devices.

# Author Contributions

O.V.P.: Conceptualization, Formal Analysis, Investigation, Methodology, Validation, Visualization, Writing – Original Draft, Writing – Review and Editing. J.P.: Formal Analysis, Investigation, Writing – Review and Editing. S.S.: Conceptualization, Methodology, Validation, Funding Acquisition, Project Administration, Resources, Supervision, Writing – Review and Editing.

# Conflict of Interest

The authors declare no conflict of interest.

# Acknowledgements

The authors would like to acknowledge Siarhei Zhuk for his help in the initial stages of the project. In addition, J. P. acknowledges funding by the SNSF (project no. 200021_196980). O.V.P. acknowledges funding by the Empa Research Commission.

# References

1. Karabalin, R. B. *et al.* Piezoelectric nanoelectromechanical resonators based on aluminum nitride thin films. *Appl. Phys. Lett.* **95**, (2009).
2. Dubois, M. A. & Muralt, P. Stress and piezoelectric properties of aluminum nitride thin films deposited onto metal electrodes by pulsed direct current reactive sputtering. *J. Appl. Phys.* **89**, 6389–6395 (2001).
3. Dubois, M. A. & Muralt, P. Properties of aluminum nitride thin films for piezoelectric transducers and microwave filter applications. *Appl. Phys. Lett.* **74**, 3032–3034 (1999).
4. Trolier-Mckinstry, S. & Muralt, P. Thin film piezoelectrics for MEMS. *J. Electroceramics* **12**, 7–17 (2004).
5. Fu, Y. Q. *et al.* Advances in piezoelectric thin films for acoustic biosensors, acoustofluidics and lab-on-chip applications. *Prog. Mater. Sci.* **89**, 31–91 (2017).
6. Ishihara, M., Li, S. J., Yumoto, H., Akashi, K. & Ide, Y. Control of preferential orientation of AlN

films prepared by the reactive sputtering method. *Thin Solid Films* **316**, 152–157 (1998).

7. Patidar, J. *et al.* Improving the crystallinity and texture of oblique-angle-deposited AlN thin films using reactive synchronized HiPIMS. *Surf. Coat. Technol.* **468**, 129719 (2023).
8. Aissa, K. A. *et al.* AlN films deposited by dc magnetron sputtering and high power impulse magnetron sputtering for SAW applications. *J. Phys. D. Appl. Phys.* **48**, 145307 (2015).
9. Trant, M. *et al.* Tunable ion flux density and its impact on AlN thin films deposited in a confocal DC magnetron sputtering system. *Surf. Coatings Technol.* **348**, 159–167 (2018).
10. Sinha, N. *et al.* Piezoelectric aluminum nitride nanoelectromechanical actuators. *Appl. Phys. Lett.* **95**, 4–7 (2009).
11. Ruffner, J. A., Clem, P. G., Tuttle, B. A., Dimos, D. & Gonzales, D. M. Effect of substrate composition on the piezoelectric response of reactively sputtered AlN thin films. *Thin Solid Films* **354**, 256–261 (1999).
12. Dadgar, A. *et al.* Sputter Epitaxy of AlN and GaN on Si(111). *Phys. Status Solidi Appl. Mater. Sci.* **220**, 1–6 (2023).
13. Hörich, F. *et al.* Demonstration of lateral epitaxial growth of AlN on Si (1 1 1) at low temperatures by pulsed reactive sputter epitaxy. *J. Cryst. Growth* **571**, 1–4 (2021).
14. Preobrajenski, A. B. *et al.* Influence of chemical interaction at the lattice-mismatched h-BN/Pt (111) interfaces on the overlayer morphology. *Phys. Rev. B - Condens. Matter Mater. Phys.* **75**, 1–8 (2007).
15. Siol, S. *et al.* Stabilization of wide band-gap p-type wurtzite MnTe thin films on amorphous substrates. *J. Mater. Chem. C* **6**, 6297–6304 (2018).
16. Sanz-Hervás, A. *et al.* Comparative study of c-axis AlN films sputtered on metallic surfaces. *Diam. Relat. Mater.* **14**, 1198–1202 (2005).
17. Nikishin, S. A. *et al.* High-quality AlN grown on Si(111) by gas-source molecular-beam epitaxy with ammonia. *Appl. Phys. Lett.* **75**, 484–486 (1999).
18. Luo, J., Wang, W., Zheng, Y., Li, X. & Li, G. AlN/nitrided sapphire and AlN/non-nitrided sapphire hetero-structures epitaxially grown by pulsed laser deposition: A comparative study. *Vacuum* **143**, 241–244 (2017).
19. Seah, M. P. Summary of ISO/TC 201 standard: VII ISO 15472 : 2001 - surface chemical analysis - x-ray photoelectron spectrometers - calibration of energy scales. *Surf. Interface Anal.* **31**, 721–723 (2001).
20. Siol, S. *et al.* Concepts for chemical state analysis at constant probing depth by lab-based XPS / HAXPES combining soft and hard X-ray sources. *Surf. Interface Anal.* 802–810 (2020) doi:10.1002/sia.6790.
21. Powell, C. J. Recommended Auger parameters for 42 elemental solids. *J. Electron Spectros. Relat. Phenomena* **185**, 1–3 (2012).
22. Tanuma, S., Powell, C. J. & Penn, D. R. Calculations of electron inelastic mean free paths. IX. Data for 41 elemental solids over the 50 eV to 30 keV range. *Surf. Interface Anal.* **43**, 689–713 (2011).
23. Gaarenstroom, S. W. & Winograd, N. Initial and final state effects in the ESCA spectra of cadmium and silver oxides. *J. Chem. Phys.* **67**, 3500–3506 (1977).
24. Wagner, C. D. Chemical shifts of Auger lines, and the Auger parameter. *Faraday Discuss. Chem. Soc.* **60**, 291–300 (1975).
25. Pélisson-Schecker, A., Hug, H. J. & Patscheider, J. Charge referencing issues in XPS of insulators as evidenced in the case of Al-Si-N thin films. *Surf. Interface Anal.* **44**, 29–36 (2012).
26. Greczynski, G., Primetzhofer, D., Lu, J. & Hultman, L. Core-level spectra and binding energies of transition metal nitrides by non-destructive x-ray photoelectron spectroscopy through capping layers.

*Appl. Surf. Sci.* **396**, 347–358 (2017).

27. Rao, C. N. R. & Ranga Rao, G. Nature of nitrogen adsorbed on transition metal surfaces as revealed by electron spectroscopy and cognate techniques. *Surf. Sci. Rep.* **13**, 223–263 (1991).
28. Fuggle, J. C. & Menzel, D. XPS, UPS and XAES studies of the adsorption of nitrogen, oxygen, and nitrogen oxides on W(110) at 300 and 100 K. *Surf. Sci.* **79**, 1–25 (1979).
29. Zhuk, S. & Siol, S. Chemical state analysis of reactively sputtered zinc vanadium nitride: The Auger parameter as a tool in materials design. *Appl. Surf. Sci.* **601**, 154172 (2022).
30. Wieczorek, A., Lai, H., Pious, J., Fu, F. & Siol, S. Resolving Oxidation States and X –site Composition of Sn Perovskites through Auger Parameter Analysis in XPS. *Adv. Mater. Interfaces* **10**, 2201828 (2023).
31. Kobayashi, A. *et al.* Epitaxial Junction of Inversion Symmetry Breaking AlN and Centrosymmetric NbN: A Polarity Control of Wide-Bandgap AlN. *ACS Appl. Electron. Mater.* (2022) doi:10.1021/acsaelm.2c01288.
32. Milyutin, E. *et al.* Sputtering of (001)AlN thin films: Control of polarity by a seed layer. *J. Vac. Sci. Technol. B, Nanotechnol. Microelectron. Mater. Process. Meas. Phenom.* **28**, L61–L63 (2010).
33. Shojiki, K., Uesugi, K., Kuboya, S. & Miyake, H. Reduction of threading dislocation densities of N-polar face-to-face annealed sputtered AlN on sapphire. *J. Cryst. Growth* **574**, 126309 (2021).
34. Zhang, Z. *et al.* Molecular beam homoepitaxy of N-polar AlN: Enabling role of aluminum-assisted surface cleaning. *Sci. Adv.* **8**, 1–7 (2022).